# Magneto-photoluminescence of charged excitons from MgZnO/ZnO heterojunctions

T. Makino[1][*], Y. Segawa[1], A. Tsukazaki[2,3], H. Saito[4], S. Takeyama[4], S. Akasaka[5], K. Nakahara[5], M. Kawasaki[1,6]

[1] Correlated Electron Research Group (CERG) and Cross-Correlated Materials Research Group (CMRG), RIKEN-Advanced Science Institute, Wako 351-0198, Japan

[2] Department of Advanced Material Science, University of Tokyo, Kashiwa 277-8561, Japan,

[3] PRESTO, Japan Science and Technology Agency, Tokyo 102-0075, Japan

[4] Institute for Solid State Physics, University of Tokyo, Kashiwa 277-8581, Japan

[5]R&D Headquarter, ROHM Co., Ltd., Kyoto 615-8585, Japan

[6]Department of Applied Physics and Quantum-Phase Electronics Center (QPEC), University of Tokyo, Tokyo 113-8656, Japan

We report on the photoluminescence (PL) properties of MgZnO/ZnO heterojunctions grown by plasma-assisted molecular-beam epitaxy. Influence of the applied magnetic field ($B$) on the radiative recombination of the two-dimensional electron gas (2DEG) is investigated up to 54 T. An increase in magnetic field in the range of $B \leq 20$ T results in a redshift in the PL. Abrupt lineshape changes in the PL spectra are observed at higher magnetic fields, in correlation with the integer quantum Hall states. We attempt to interpret these features using the conventional model for the 2DEG-related PL based on the transition between the 2DEG and a hole as well as a model taking a bound state effect into account, i.e., a charged exciton. The comparison about the adequateness of these models was made, being in favor of the charged exciton model.

## I. INTRODUCTION

Optical spectroscopy has been adopted for studying the energy spectrum of electronic states of two-dimensional electron gas (2DEG) in low-dimensional heterostructures with the aim of probing the states at energies away from the Fermi level [1,2]. The energy spectrum of the 2DEG is perturbed by the photoexcited holes via electron-hole Coulomb interaction. This holds true even in GaAs heterostructures although the Coulomb interaction is weak. Coulomb interaction may play major role for ZnSe heterostructures because ZnSe has an exciton with a 20 meV binding energy, compared with 4 meV for GaAs. Keller *et al.* [3] reported the modification of the optical spectra with this interaction for ZnSe modulation-doped quantum wells (QWs), the phenomena of which are exemplified as exciton-like resonance in high magnetic fields and combined electron-exciton processes at small filling factors. The Coulomb interaction is expected to be even more important in another II-VI semiconductor—ZnO (the exciton binding energy in bulk ZnO is 60 meV). Recently, quantum Hall effect has been observed for *n*-type charged carriers in ZnO/Mg$_x$Zn$_{1-x}$O single heterojunctions (SHJs) [4-6]. Here, a mismatch in the spontaneous polarization between MgZnO and ZnO causes a strong electric field to accumulate two-dimensional electron gas (2DEG) in a triangular-shaped potential near the interface of a ZnO/Mg$_x$Zn$_{1-x}$O SHJ. Further efforts in recent years resulted in the observation of the fractional quantum Hall effect, quantum Hall insulating phase, the enhancement of the effective Landé factor, and the increase in the mobility over 700,000 cm$^2$V$^{-1}$s$^{-1}$ [7-10]. Despite recent observation of the 2DEG-related photoluminescence (PL) in the absence of magnetic field [11], unlike the magneto-transport measurements routinely used to characterize the 2DEG, magneto-PL data have not yet been reported for ZnO HJs, to the best of our knowledge.

In this paper, we present optical studies of MgZnO/ZnO SHJs, the purpose of which is to elucidate the effect of strong Coulomb interaction on the optical spectra related with 2DEG. The 2DEG-related optical transitions in low-temperature PL spectra are observed. Abrupt lineshape changes in the PL lineshape at strong magnetic fields are correlated with the integer quantum Hall states corresponding to the small filling factors such as $\nu = 2$ and $\nu = 3$. We attempt to interpret these phenomena with models based on the charged exciton and on the two-dimensional electron-hole (2De-h) transition.EuO films were deposited on YAlO₃ substrate using a pulsed laser deposition system with a base pressure lower than $8 \times 10^{-10}$ Torr[22].

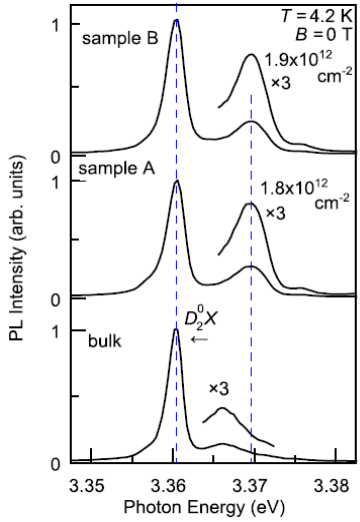

FIG. 1. (Color online) Low-temperature PL spectra ($T = 4.2$ K and $B = 0$ T) in single heterojunctions (SHJs) and ZnO bulk crystal. The sample A is Mg$_{0.11}$Zn$_{0.89}$O/ZnO SHJ, while the sample B is Mg$_{0.14}$Zn$_{0.86}$O/ZnO SHJ. The emission about 6 meV below the exciton resonance ($X_0$) for SHJ is attributed to recombination involving electrons from the 2DEG at ZnO interface ($X^-$).

## II. EXPERIMENTAL PROCEDURES

We performed magneto-PL measurements under pulsed magnetic fields for two Mg$_x$Zn$_{1-x}$O/ZnO ($x = 0.11$ and 0.14) SHJs with the growth axis (crystallographic $c$ axis) parallel to the applied magnetic field ($\boldsymbol{B} \parallel c$) as well as for a ZnO bulk crystal (reference sample). A 325-nm line from a continuous-wave He-Cd laser was used as the excitation source. The wavevector $\boldsymbol{k}$ of the unpolarized exciting photons is parallel to the crystallographic $c$ axis ($\boldsymbol{k} \parallel c$). An optical fiber, mounted in a He cryostat at a temperature ($T$) of 4.2 K, was used for both photoexcitation and PL collection under backward scattering geometry [12]. The optical fiber used covers the spectral wavelength range from 200 to 800 nm. The core diameter of the fiber is 200 μm. The spectral resolution of the monochromator for PL collection without the polarization analysis is approximately 0.5 nm. The pulsed magnetic field was generated by capacitor discharge. The pulse width is 37 ms and the maximum field is 54 T. Readers should refer to Refs. [13] and [14] for the details of the measurement system. Two SHJ samples were fabricated by plasma-assisted molecular-beam epitaxy (MBE) [15]; one is called sample A with 100-nm-thick undoped buffer layer followed by a deposition of 540-nm-thick Mg$_{0.11}$Zn$_{0.89}$O layer, the other is called sample B with 200-nm-thick undoped buffer layer followed by 180-nm-thick Mg$_{0.14}$Zn$_{0.86}$O layer [16]. The carrier density ($n_e$) and mobility of the 2DEG in sample A (sample B) were evaluated through the electronic transport (Hall effect) measurements to be $1.8 \times 10^{12}$ cm$^{-2}$ and 7000 cm$^2$V$^{-1}$s$^{-1}$ ($1.9 \times 10^{12}$ cm$^{-2}$ and 6000 cm$^2$V$^{-1}$s$^{-1}$ for sample B) [17]. The bulk crystal is a commercial single crystal from Tokyo-Denpa Company Ltd., grown by hydrothermal method, followed by chemo-mechanical polish (Mitsubishi Chemicals) [18,19].

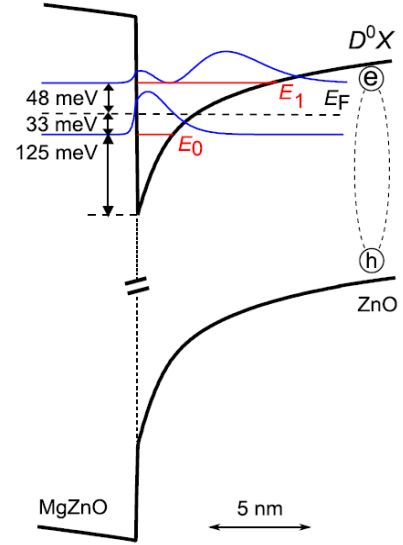

FIG. 2. (Color online) The band potential at the MgZnO/ZnO heterointerface. The energy levels are obtained from a self-consistent calculation. Two electron levels are obtained with energies of 158 and 206 meV above the notch potential. Also shown are the electron probability functions for the two occupied levels. Schematic for the recombination process of bound exciton ($D^0X$) is illustrated.

## III. RESULTS AND DISCUSSION

PL spectra measured at 4.2 K for the SHJ (samples A and B) and ZnO bulk crystal are shown in Fig. 1. In the absence of a magnetic field, the spectra for both samples are dominated by donor bound exciton emissions ($D^0_2X$), typically originating from the ZnO buffer layer or from the near-surface region of the bulk crystal. The recombination energy of $D^0_2X$ emission is 3.361 eV [20]. As schematically drawn in Fig. 2, photocreated holes are diffused away from the 2DEG channel by the electric field, and excitons are formed in the buffer layer or substrate, which leads to the observation of the bound exciton ($D^0X$)

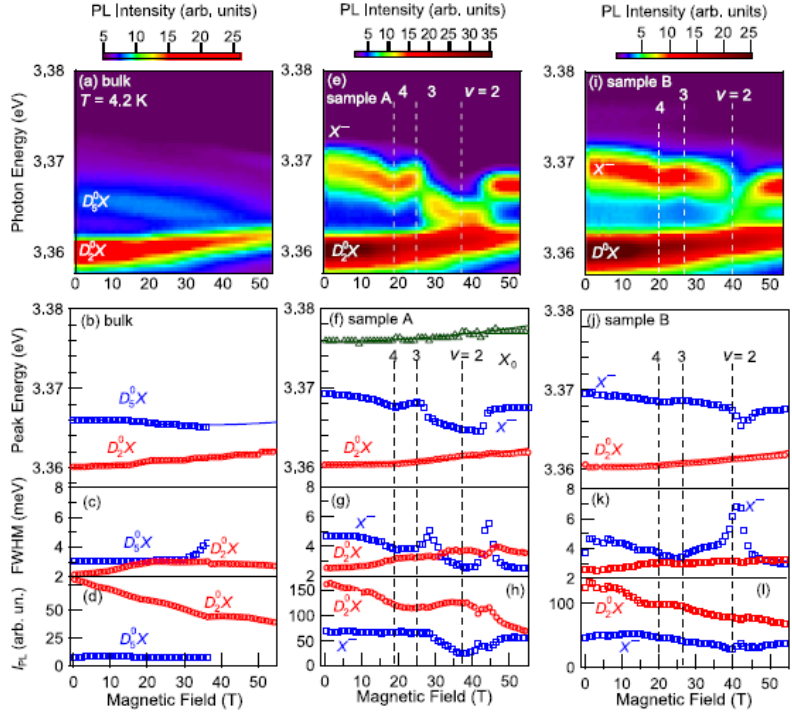

FIG. 3. (Color online) Photoluminescence energy and intensity (color scale) of the magneto-PL as a function of the external magnetic field for (a) the bulk, (e) the SHJ sample A, and (i) sample B. The integer filling factors ($v$) are indicated with vertical broken lines. Magnetic-field dependences of the characteristic quantities for $X_0$ (triangles), $D^0 X$ (circles), and $X^-$ (squares) peaks are also shown: (b),(f),(j) peak energies; (c),(g),(k) widths; and (d),(h),(l) integrated intensities. The solid curves in (b), (f), and (j) for $D^0_5 X$, $D^0_2 X$, $X_0$, and $X^-$ show the results of fit using Eq. (1).

PL [21,22]. A weaker PL peak observed for the bulk is assigned to the donor-bound exciton ($D^0_5 X$) [20]. We found a PL peak ($X^-$) in the HJ, the energy of which is different from that of $D^0_5 X$, present in the bulk ZnO spectrum. The notations of $D^0_2 X$ and $D^0_5 X$ are adopted from Ref. [20] (cf. Fig. 52) by identifying from their energetic values.

The nature of the $X^-$ PL is different from that of the $D^0_5 X$ emission, which is clearly seen from the magnetic-field dependent data. Figures 3(a), 3(e), and 3(i) show the magneto-PL for the bulk and the SHJs (samples A and B) in which the emission intensities are shown on a color scale. The $D^0_5 X$ peak disappears at around 35 T, while the $X^-$ PL is seen at all the accessible magnetic fields ($B \leq 54$ T), accompanying sudden jumps in its energy near the filling factors $v = 2$ and 3. The definition of the filling factor is $v = n_e h/eB$, where $n_e$ is the carrier density and the rest of the symbols have their standard meanings. The external magnetic field strengths for which the $v$'s take integer values are shown as vertical broken lines. Figure 3(b) shows the peak energies of two bound excitons ($D^0_5 X$ and $D^0_2 X$), while Figs. 3(f) and 3(j) show those for the free exciton ($X_0$), the 2DEG-related emission

($X^-$), and the bound exciton ($D^0_2 X$) observed in the SHJs. Based on the correlation with the integer quantum Hall states, the PL peak, $X^-$, is more likely to be assigned to the 2DEG-related recombination occurred at the interface. This assignment is consistent with the recent observation by Chen *et al* [11]. The two-dimensional electron gas related PL is observed at the lower energy side of the main $D^0_2 X$ peak in ZnO/Mg$_{0.2}$Zn$_{0.8}$O SHJ grown by metal-organic chemical vapor deposition, which is in contrast to our observation. Our peak energy is higher than that of $D^0_2 X$ peak. Although the reason is not clear at the moment, stronger localization effect in ZnO/Mg$_{0.2}$Zn$_{0.8}$O SHJ may be responsible for this difference. It should be noted that three kinds of the 2DEG-related recombination have been reported so far: (1) recombination between the two-dimensional electrons and the localized hole near the interface [11,23], (2) two-dimensional electron-hole recombination between the 2DEG and a photoexcited hole (2De-h) [21], and (3) recombination of the charged excitons [24].

Having established the relationship of the $X^-$ peak to the 2DEG at the interface, we further discuss its detailed assignment. As earlier-mentioned, there exist three candidates for the assignment of the 2DEG-related PL. First, we argue about the recombination between the 2DEG and a localized hole near the interface. We can rule out this possibility by the following reason. The larger the spatial overlap of the electron-hole wavefunctions is, the higher the recombination probability is expected to become. Therefore, we limit ourselves to the case of radiative recombination with localized holes near the interface. Figure 2 shows the result of a self-consistent calculation of the energy band potential and energy levels at the heterointerface for ZnO SHJ (sample A) near the interface. The lineshape of the energy band potential suggests that the trapping efficiency for the holes generated initially in the alloyed barrier (i.e., MgZnO) is expected to be higher than those created in ZnO layer. The holes in ZnO tend to diffuse away from the 2DEG channel. Such a shallow interfacial centers at the barrier layer side is indeed known to be involved in the 2DEG-related recombination process in AlGaN/GaN heterojunction [23], the final state of the transition is located between the valence bands of alloyed AlGaN and GaN. Similar PL has been previously reported in ZnO heterojunction [11]. If we recall that the energy distance between lowest quantized state ($E_0$) and the bottom of the potential notch exceeds 100 meV in our case, the energy of such a transition must be higher than the band gap energies of ZnO, which is in contradiction with our experimental observation. In our case, the energy of the $X^-$ peak is lower than the exciton resonance energy.

Next we argue about the direct recombination between the 2DEG and a photocreated hole. In this case, the hole tends to diffuse away from the 2DEG channel, which seems to give rise to a redshift of the PL. This consistently explains the energetic position of the $X^-$ peak. According to previous reports for other semiconductors [3], the PL band at 0 T tends to be rather broad corresponding to the 2DEG Fermi energy. The FWHM of the $X^-$ in our case is much smaller as shown in Figs. 3(g) and 3(k). Nevertheless, we cannot rule out the possibility of this assignment because indirect character of the transition from the Fermi level can weaken its high energy component. In other words, the lineshape of the PL does not necessarily reflect that of the density of occupied electronic states in the 2DEG. In spite of high enough resolution in optical measurement system, we did not detect the splitting of the PL. Typically, the PL band starts to split into a set of PL lines at relatively weak magnetic fields typically corresponding to $\nu \approx 6$ [2,3]. Even in case that the weakening effect of high energy component is dominant, the transitions from the bottoms of the respective energy states due to the Laudau level splitting. Based on the missing splitting, it is not very reasonable to assign this peak to the direct recombination between the 2DEG and a hole.

On the other hand, due to the bound state effect, there is no report of the splitting of the charged exciton PL related to the Landau level splitting. Therefore, the $X^-$ PL peak is more likely to be assigned to the charged exciton. At 0 T, the $X^-$ PL band is located about 6 meV below the free-exciton emission ($X_0$) observed at 3.375 eV. This Stokes shift is similar to the binding energy of charged exciton in modulation-doped ZnO QWs [25]. When we focus ourselves on the magnetic field dependence of the peak energy in the range of $B < 20$ T, both the samples A and B exhibit the redshift with an increase in the magnetic field as shown in Figs. 3(e) and 3(f) (open squares). Its behavior is significantly different from the shift of the bulk excitons ($D_2{}^0X$ and $X_0$). By accounting for the fact that lower energy component of the split Zeeman term ($n_\downarrow$ and $n_\uparrow$) contributes predominantly, the magnetic field dependence of the peak energy $E$ can be described by the following expression [26].

$$E = E_0 + \frac{1}{2} g^* \mu_B B \left( \frac{n_\uparrow - n_\downarrow}{n_\uparrow + n_\downarrow} \right) + DB^2,$$

(1)

where $E_0$ is the PL energy at 0 T, $g^*$ an effective Landé factor, $\mu_B$ Bohr magneton, and $D$ a quadratic diamagnetic coefficient. The ratio $n_\uparrow/n_\downarrow$ is given by a Boltzmann distribution. Because the singlet charged excitons are rather stable at weak magnetic field according to Hill's theorem[27], the Zeeman part of magnetic field dependence of the peak energy is determined by $g^* = |g_{e-}g_h|$ ($\approx 0.8$). Assuming the diamagnetic shift coefficient $D$ being equal to the literature value [28-30], the result of calculation using Eq. (1) yielded in a solid curve shown in Fig. 3(f) for $X^-$. This is not in a good agreement with an experimental result. It is probable to attribute this redshift to an enhancement of the binding energies of the charged exciton. The enhancement of the binding energy has so far been reported in a variety of QWs [31]. On the other hand, the field-induced evolution of the bound exciton energies is explained in terms of the dominance of the Zeeman term over the counterpart—the quadratic diamagnetic term.

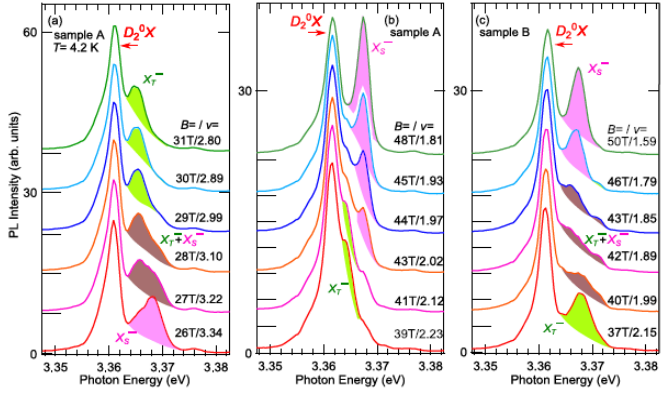

FIG. 4. (Color online) Spectra measured at different magnetic fields corresponding to various filling factors $\nu$ as specified for each spectrum: (a) 24 T $\leq B \leq$ 30 T for sample A, (b) 37 T $\leq B \leq$ 45 T for sample A, and (c) 37 T $\leq B \leq$ 50 T for sample B. The singlet and triplet charged excitons ($X_S^-$ and $X_T^-$) are depicted by pink (medium gray) and green (thick gray) shadings.

The results of fits for both $D_5^0X$ and $D_2^0X$ are in reasonably good agreement with the experimental data. The overall dependence is determined by the relative magnitudes of the linear Zeeman and quadratic diamagnetic terms, which leads to the different behavior of the field-induced peak shift between $D_2^0X$ and $D_5^0X$. It should be noted that an effective Landé $g^*$ factor for $D_5^0X$ can be different from that of neither $X_0$ nor $D_2^0X$ due to mixing of the quasidegenerate valence states by the defect potential. With assumption of the quadratic diamagnetic coefficient $D$ being equal to the literature value , the effective Landé value for $D_5^0X$ ($D_2^0X$) is evaluated to be $g^* \approx 2$ (0) through the least-square fit using Eq. (1). We did not observe the splitting induced by the application of the magnetic field for $D_5^0X$ and $D_2^0X$ PL because of the vanishing $g$ value in the case of $D_2^0X$ and of the quenching

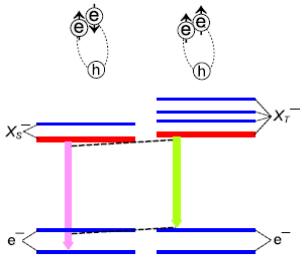

FIG. 5. (Color online) Schematic drawings for $X_S^-$ and $X_T^-$ and explanation of the observed PL energies in terms of the spin states of the final states after recombination. The triplet PL energy remains lower than that of the singlet, as indicated by the lengths of the arrows, although the energy position of the triplet state is higher.

of the $D_5^0X$ PL at $B = 35$ T.

If we try to interpret the redshift in the relatively weak field region with the direct recombination from the 2DEG to the hole (2De-h), there is a previous work reporting the Fermi level energy which is approximately independent of the magnetic field below $B \leq 10$ T in modulation-doped ZnSe QW [3]. This is not, however, true in our case because the indirect character of the transition rules out the possibility that the PL peak corresponds to the Fermi energy. If the Zeeman effects on the energies of the quantized electronic states and the valence band are extraordinarily strong so as to cancel the blueshifting behavior associated with the $B$-induced formation of the Laudau levels, one may be able to explain this redshift. Such an extraordinary large $g$ value has not been reported in ZnO SHJ containing the high-density 2DEG ($n_e > 10^{12}$ cm$^{-2}$) [9,10,25]. This observation is decisive to the assignment to the charged exciton despite the high electron concentration in our SHJs. If we are allowed to take the Coulomb effect on the 2De-h into account, the redshift can be attributed to the $B$-induced binding energy enhancement of the bound state composed of the 2DEG and a photocreated hole. The existence of such a bound state, called a Mahan exciton, has recently been found experimentally [32] and confirmed theoretically [33] in $n$-doped ZnO films. In our view, the Mahan exciton is more or less a kind of "many-body" extension of the charged exciton which involves only one electron from the 2DEG channel. The problem as to how many electrons are involved for the bound states is beyond the main scope of our work which is the Coulomb modification of the 2DEG-related optical spectra as already stated in the introduction.

Next, the discussion is made for the peak energy shift between $\nu \approx 4$ and 3. Magnetic-field-induced blueshift is observed for both the samples A and B. A reasonable interpretation relying on the model of the charged exciton is attribution of this blueshift to the diamagnetic shift. On the other hand, the blueshift and energy jumps near the filling factors can be easily explained in the 2De-h picture in terms of the magnetic field dependence of the energies of the Laudau level, the valence band, and the Fermi level. Simple calculation of the Fermi energy as a function of magnetic field [34] did not give agreement with the experimental peak energy shift. It is expected that execution of elaborate calculations taking the many-body effects into account results in better agreement, which is, however, beyond the intended scope of this paper. There are many numerical factors which must be taken into account to quantitatively model the magneto-PL energies. Among them are magnetic field dependences of the band gap renormalization, conduction and valence band energies, binding energy of the charged exciton, and energies of the diffused hole.

We try to interpret the magnetic field dependence of the FWHM of PL from the viewpoint of the adequateness of the assignment to both postulations (the charged exciton and the 2De-h). Figures 3(g) and 3(k) respectively show the FWHMs as a function of the magnetic field for the samples A and B. As long as a single peak is assumed in the width evaluation, the $X^-$ width increases with thresholds near the filling factors $\nu \approx 3$ and 2. As shown in Fig. 4, a closer look at the spectral features for sample A in these magnetic field regions (26 to 29 T) reveals a splitting into two different components, which results in the seeming enhancements in the FWHM. As schematically drawn in Fig. 5, the splitting observed at strong magnetic field is known to be originated from the singlet $(X_S^-)$ -triplet $(X_T^-)$ splitting in the picture of charged exciton. Attempts will be made hereafter as to how we can soundly interpret these experimental findings based on the assumption of the assignment to the charged exciton. At zero magnetic field, $X_S^-$ is preferred [27,35], while $X_T^-$ should be the ground state at the extreme magnetic field limit [35,36]. As shown in Fig. 4(a), a natural interpretation drawn from this $B$-dependent stability is that the PL intensity in sample A is redistributed significantly in favor of $X_T^-$ at 29 T ($\nu = 2.57$) after the coexistence with $X_S^-$. Then, the PL of $X_T^-$ weakens at fields corresponding to $\nu \approx 2$ [lowest trace in Fig. 4(b)]. At 41 T ($\nu = 1.82$) [Fig. 4(b)], however, the intensity of $X_S^-$ starts to become larger than that of $X_T^-$ again. This is also true even at the maximum available field (i.e., $B = 54$ T). Based on this, the singlet state is probably the ground state of charged exciton for the accessible field strength range ($0 \leq B \leq 54$ T). As drawn schematically with the red bold lines in Fig. 5, the *resonance* energy of $X_T^-$ is very close to that of $X_S^-$ at $B \geq 26$ T. The energy of $X_T^-$ is, however, still higher than that of $X_S^-$ even at $B = 54$ T. We observe $X_T^-$ PL peak that has higher intensity over intermediate magnetic field range near the filling factor of $\nu = 3$. This could be due to the fact that the spin polarization ($P$) of the 2DEG at $\nu = 3$ is $P = 1/3$, which is higher than that at $\nu = 2$ ($P = 0$). The higher the spin polarization is, the more probably the spins of electrons in a triplet charged exciton are oriented in a same direction. The $X_T^-$ PL energy is lower than the $X_S^-$ PL energy despite the absence of a singlet-triplet crossover of a charged exciton ground state. As shown schematically in Fig. 4(a), these seemingly incompatible results may be explained by considering that the spin state of electrons which remain after $X_T^-$ recombination is different from that after $X_S^-$ recombination [37]. If the final state of $X_T^-$ recombination resides the upper state unlike the case of $X_S^-$ recombination, $X_T^-$ PL energy in Fig. 5 is larger than that of $X_S^-$ PL, as indicated by the lengths of the arrows. As drawn with the red bold lines, because the resonance energies are closely situated, the splitting energy of the final state (i.e., the electron Zeeman energy $\Delta E_e = g_e \mu_B B$) determines the difference in the PL peak energies (i.e., the arrow lengths) [37]. The assignment of $X_T^-$ consistently explains the magnetic field dependence of the $X^-$ PL intensity. As shown in Figs. 3(h) and 3(l), the PL intensity of $X^-$ starts to be decreased with an increase in the field over the range where the triplet charged exciton ($X_T^-$) is dominant because the triplet charged exciton is optically dark [35,37].

If the PL band is assigned to the direct recombination between the 2DEG and a photocreated hole, inspection of the existing literature can explain the above-mentioned splitting correlated with the small filling factors in terms of the effect of the resonant many-body interaction with a continuum of spin excitation of the 2DEG, involving spin-flip and Auger processes [38,39]. In the case of modulation-doped CdTe QW, such a splitting is reported to start from $\nu = 3$ and disappear at $\nu = 2$. The splitting energy is approximately 3 meV [39], which is comparable with that in ZnO SHJs. The appearance of the splitting in our samples is quite similar to the observation in CdTe QW. Further experimental study such as polarized magneto-PL is necessary in order to draw more conclusive assignment for this 2DEG-related luminescence band.

We discuss the effect associated with the ternary alloy character of the cap layers. A comparison of the PL width at 0 T is made for the samples having low and high Mg concentrations. The $Mg_{0.11}Zn_{0.89}O$ SHJ yielded a value of 5 meV, while $Mg_{0.04}Zn_{0.96}O$ SHJ results in 4 meV. This suggests that the alloy character has a minor influence on the broadening of $X^-$ peak, compared to the case of Mg concentration dependence of the localized excitonic PL width. It is considered that the two-dimensional electron less sensitively feels the effect of concentration fluctuation in SHJs having very high mobility than in the case of the localized exciton, giving rise to smaller broadening effect on the Fermi energy.

## IV. SUMMARY

We have assessed optical properties of $Mg_{0.11}Zn_{0.89}O/ZnO$ and $Mg_{0.14}Zn_{0.86}O/ZnO$ SHJs to investigate the modification of spectra in structures with strong Coulomb interaction. We confirmed the existence of 2DEG-related PL peak in the SHJs, which was reported by Chen *et al*. The application of a magnetic field corresponding to the filling factors $v = 2$ and 3 yielded observation of pronounced PL anomalies as exemplified with energy jumps and splittings. Comparison of the soundness of the spectral interpretation favors the assignment to the charged exciton ($X^-$) for explaining our findings such as the redshift at $B < 20$ T and splitting at $v = 2$ and 3.

Acknowledgements—This research was supported by the Japan Society for the Promotion of Science (JSPS) through the grant "Funding Program for World-Leading Innovative R&D on Science and Technology (FIRST Program)", initiated by the Council for Science and Technology Policy (CSTP) and in part supported by KAKENHI (Grant Nos. 21104502 and 24540337), Japan (T. M.). We wish to thank Dr. Y. Kozuka for his help in measuring the electron transport properties.

* tmakino@riken.jp